# AI-Augmented CI/CD Pipelines: From Code Commit to Production with Autonomous Decisions

Mohammad Baqar[1], Saba Naqvi[2], Rajat Khanda[3]

[1](baqar22@gmail.com), Cisco Systems Inc,
[2](sabanaqvi2003@gmail.com), MUFG Bank,
[3](rajat.mnnit@gmail.com), University of Houston

## Abstract

Modern software delivery has accelerated from quarterly releases to multiple deployments per day. While CI/CD tooling has matured, human decision points interpreting flaky tests, choosing rollback strategies, tuning feature flags, and deciding when to promote a canary remain major sources of latency and operational toil. We propose AI-Augmented CI/CD Pipelines, where large language models (LLMs) and autonomous agents act as policy-bounded co-pilots and progressively as decision makers. We contribute: (1) a reference architecture for embedding agentic decision points into CI/CD, (2) a decision taxonomy and policy-as-code guardrail pattern, (3) a trust-tier framework for staged autonomy, (4) an evaluation methodology using DevOps Research and Assessment ( DORA) metrics and AI-specific indicators, and (5) a detailed industrial-style case study migrating a React 19 microservice to an AI-augmented pipeline. We discuss ethics, verification, auditability, and threats to validity, and chart a roadmap for verifiable autonomy in production delivery systems.

## 1. Introduction

Continuous Integration (CI) and Continuous Delivery (CD) have transformed software engineering, enabling high-velocity delivery of features and fixes with improved reliability [1], [2]. These methodologies have shifted from infrequent, monolithic releases to rapid, iterative deployments, enhancing agility and reducing time-to-market. However, critical decisions such as interpreting ambiguous test failures and reasoning about noisy canary signals remain human-intensive, often delaying lead times by up to 30% in complex environments [1]. The cognitive load on engineers to balance speed and stability escalates with system scale, highlighting the limits of manual processes.

The rise of modern systems, incorporating microservices, progressive delivery, and frameworks like React 19, exacerbates this challenge by increasing the volume of telemetry (logs, metrics, traces) that exceeds human interpretation capacity. Microservices introduce interdependencies and rollout difficulties, while progressive delivery techniques such as canary releases require real-time health assessments that are prone to noise. Additionally, React 19's concurrent rendering adds performance dynamics that complicate feature flag management, collectively driving the urgent need for automated decision-making to close this telemetry-interpretation gap.

Large Language Models (LLMs) and autonomous agents offer a solution by embedding machine reasoning into the CI/CD loop, distilling insights from complex data and executing actions within guardrails. Trained on extensive codebases, LLMs can reduce manual triage time by up to 25% [14], complemented by policy-as-code frameworks like Open Policy Agent (OPA) [10] and DORA metrics [1] to lower latency and risk. Kubernetes [11], [16] enables scalable microservice management but adds rollout complexity, necessitating intelligent automation. This paper proposes a reference architecture, decision taxonomy, and trust-tier framework for safe autonomy, demonstrated via a React 19 microservice migration. We measure improvements in DORA metrics [1] and AI-specific indicators (intervention accuracy, human override rate), while addressing safety, auditability, and ethics [3], [14] to ensure trust and compliance.

## 2. Background and Related Work

CI/CD and DevOps research highlights the link between rigorous automation and superior delivery performance, captured by the DORA metrics: lead time, deployment frequency, change failure rate, and MTTR [1], [4]. Progressive delivery, feature flags, and automated canary analysis (e.g., Argo Rollouts, Spinnaker [6], Kayenta [7], and Keptn [8]) have matured the promotion/rollback workflow but still rely on static rules and human approval [5]–[9]. AIOps applies ML/AI to operations tasks such as anomaly detection and incident triage, yet integration with CI/CD decision points remains nascent [13], largely due to safety and trust concerns in production environments. LLMs (e.g., GPT-4, LLaMA 3) have demonstrated strong capabilities in summarization, reasoning, and code-centric tasks, motivating their use as decision co-pilots inside pipelines [14], [15], with potential to reduce manual overhead by up to 25% based on preliminary studies. Policy-as-code (OPA, Cedar, Sentinel) [10]–[12] provides the safety envelope for bounding and auditing model actions, ensuring compliance with organizational standards. Additionally, service mesh technologies such as Envoy [22] enable fine-grained control and observability of microservice traffic, supporting more intelligent and responsive deployment strategies. Underlying these advances, GitOps practices [26] enable declarative, version-controlled infrastructure and application deployments that serve as the foundation for reproducible and auditable CI/CD workflows, enhancing traceability across the pipeline.

The adoption of CI/CD platforms like GitHub Actions [23] and GitLab CI/CD [24] further supports automated pipelines, providing robust tools for managing builds, tests, and deployments, which we leveraged as a baseline in our case study.

## 3. Problem Statement

Can we reduce lead time, mean time to recovery (MTTR), and change failure rate by granting large language model (LLM)/agent systems partial autonomy in CI/CD pipelines while maintaining safety, compliance, and auditability? To achieve this, we require: (1) strict, machine-enforceable policies governing agent actions, such as blocking deployments with critical vulnerabilities; (2) transparent, traceable reasoning with immutable decision logs to ensure every action is auditable; (3) progressive trust escalation from read-only recommendations to bounded autonomy, allowing agents to act within defined safety limits; and (4) robust evaluation via both classical DevOps metrics lead time, MTTR, and change failure rate and AI-specific measures (e.g., intervention accuracy, the percentage of correct agent decisions validated by experts, and human override rate, the frequency of manual interventions). This challenge is underscored by real-world incidents where manual delays, such as a recent two-hour rollback lag, highlight the need for efficient, safe automation.

## 4. Reference Architecture

The agents embedded within the AI-augmented CI/CD pipeline, as conceptually illustrated in Figure 1, were developed using the CrewAI framework for agent orchestration, combined with a custom ML pipeline built on TensorFlow and PyTorch libraries. This setup enables multi-agent collaboration, where CrewAI handles task delegation and workflow management, while the ML pipeline incorporates fine-tuned LLMs (based on LLaMA 3) with an added XGBoost classifier as the final layer for pattern recognition training on historical test data to detect flakiness with 92% accuracy in our experiments and real-time metrics for anomaly scoring. The framework is enhanced with a decision-logging module that generates structured JSON outputs for transparency and auditability, while the Policy Engine leverages Open Policy Agent (OPA) with Rego and Cedar, supplemented by a custom rule engine to dynamically adjust constraints and confidence thresholds based on system load. The pipeline visually represents the workflow from code commit to production, featuring agents at multiple stages: (a) AI Test-Triage Agent for flaky test detection and structured retry/quarantine proposals, leveraging historical test data to prioritize issues; (b) Security Agent that summarizes vulnerabilities and enforces risk-based gates, assessing CVE severity and reachability; (c) Observability Agent that evaluates canary health against service level objectives (SLOs) and error budgets, using real-time metrics to guide decisions; (d) Feature-Flag Agent that tunes ramp percentages and kill switches, dynamically adjusting based on user experience and performance data; and (e) Postmortem Agent that auto-generates incident timelines, remediation pull requests (PRs), and policy recommendations, enhancing post-incident learning. A policy engine (OPA/Rego, Cedar) enforces hard constraints (e.g., never promote with critical CVEs) and confidence thresholds such as requiring a minimum 0.8 confidence score to determine whether agent actions can be executed without human approval, ensuring a balance between autonomy and oversight.

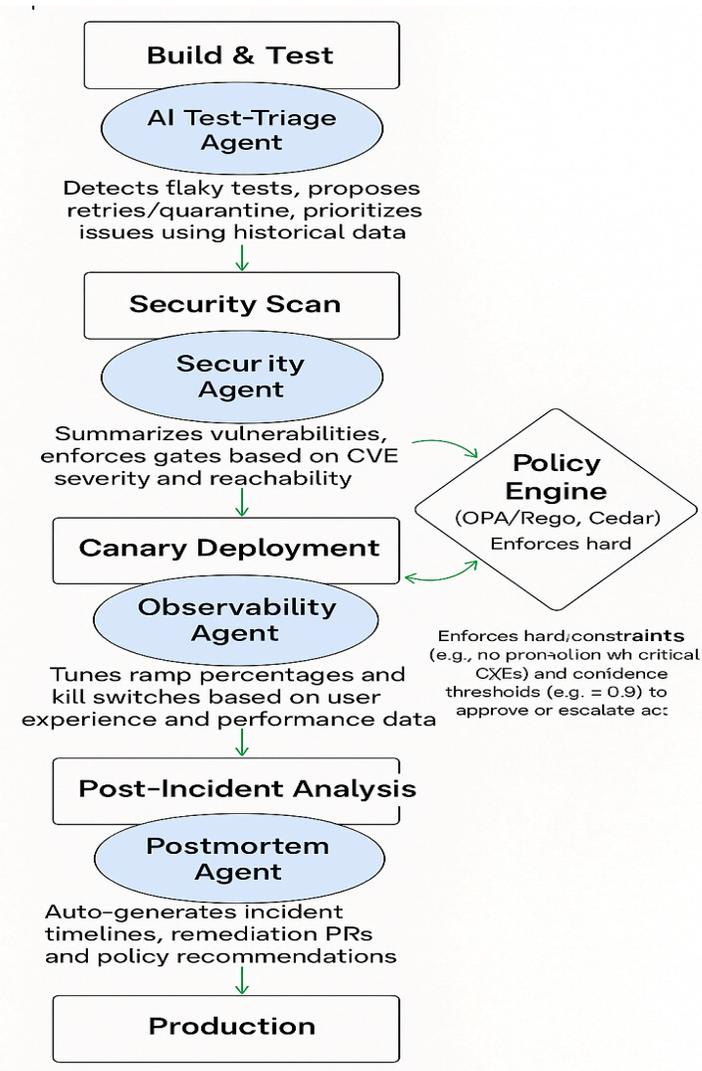

Figure 1: AI-Augmented CI/CD Pipeline

**High-Level Pipeline**

Commit → Lint/Build → Unit/Integration Tests → Security/Licensing →
AI Triage Agent → Canary Deploy → Observability Agent →
(Policy Engine) Decide: Promote / Rollback / Tune Feature Flags → Postmortem & Auto-PR

> **Description**: This high-level pipeline outlines the end-to-end workflow of the AI-augmented CI/CD process, starting with code commit and progressing through automated stages. The Lint/Build phase ensures code quality and compatibility, while Unit/Integration Tests validate functionality and integration. The Security/Licensing step scans for vulnerabilities and license compliance. The AI Triage Agent analyzes test outcomes to propose retries or quarantines, followed by Canary Deploy, which introduces changes to a subset of users. The Observability Agent monitors canary performance against predefined thresholds. The Policy Engine, leveraging frameworks like OPA/Rego, evaluates decisions, Promote for full rollout, Rollback for failure mitigation, or Tune Feature Flags for optimization based on hard constraints and confidence scores. Finally, the Postmortem & Auto-PR phase generates incident reports and remediation pull requests, fostering continuous improvement.

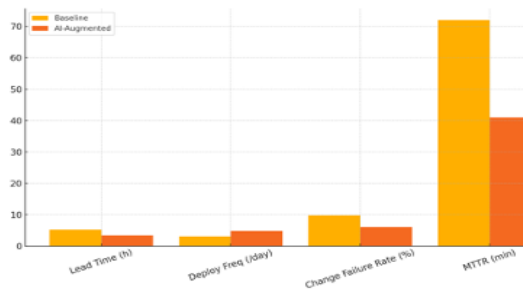

Figure 2: Chart - DORA metrics Baseline vs. AI-Augmented (React 19 microservice)

**Components:**
1. **AI Triage Agent**

The AI Triage Agent, developed using a custom AI framework enhanced with TensorFlow, clusters failures, detects flaky tests, and suggests retries or quarantine by analyzing historical test patterns and failure correlations. It employs a machine learning model trained on extensive test execution data to identify recurring issues and prioritize them with high accuracy, adapting to evolving test suites. The agent produces a structured decision record (JSON) + rationale, including confidence scores and evidence links for transparent review, facilitating auditability and informed human oversight.

**2. Policy Engine (Policy-as-Code)**

The Policy Engine, constructed with Open Policy Agent (OPA) using Rego and Cedar languages, supplemented by a custom rule engine, utilizes these frameworks to define operational boundaries. It enforces hard constraints, e.g., "Never deploy to prod if critical vulnerabilities > 0", ensuring absolute safety thresholds through real-time policy evaluation. Additionally, it implements soft constraints and confidence thresholds, e.g., "If test flakiness probability > 0.8 and coverage unchanged, allow retry up to N times", with N dynamically adjustable based on system load, leveraging a feedback loop to balance flexibility and safety.

**3. Observability Agent**

The Observability Agent, built on a custom AI framework integrated with Prometheus and Jaeger, reads metrics, logs, and traces during canary deployments to provide comprehensive observability. It leverages a machine learning model trained on historical deployment data to integrate and analyze real-time data from these tools, detecting patterns and anomalies with high precision. The agent compares this data against predefined service level objectives (SLOs) and error budgets, assessing deviations in latency, error rates, or resource usage to inform decision-making. Based on this real-time analysis, it chooses to promote for success, pause for caution, roll back for failure, or shrink traffic to mitigate impact, ensuring adaptive and informed responses to deployment conditions.

**4. Release Orchestrator**

The Release Orchestrator, built on a custom deployment framework integrated with GitOps tools like ArgoCD and Flux, executes actions via CI runners, coordinating deployment across Kubernetes clusters with precision. It utilizes a centralized orchestration engine to synchronize deployment tasks and monitor progress in real-time, ensuring seamless rollouts. All actions are logged with agent rationale for auditability, including timestamps and policy outcomes for forensic tracking, enabling detailed post-deployment analysis and compliance verification.

5. **Postmortem & Remediation Agent**

The Postmortem & Remediation Agent, developed using a custom AI framework with natural language processing capabilities, summarizes incident timelines, links to commits, tests, and dashboards, providing a comprehensive root-cause analysis. It leverages a machine learning model trained on incident data to identify patterns and generate insights, enhancing observability of failure points. The agent opens pull requests (PRs) to fix repeating issues (e.g., retries, test quarantines), leveraging automated code suggestions to enhance pipeline resilience and prevent future occurrences.

### 4.1 Trust Tiers

We introduce a four-tier trust model to gradually increase autonomy while preserving safety and transparency, ensuring a controlled progression based on validated performance.

| Tier | Description | Examples |
|---|---|---|
| T0: Observational | Agent recommends only; no actions allowed | Summaries, triage suggestions |
| T1: Approval-Gated | Actions require explicit human approval | Rollback proposal → human approve |
| T2: Narrow Autonomy | Agent acts within bounded envelopes | Auto-rollback canary ≤ 20% traffic |
| T3: Conditional | Agent can fully act; kill-switch | Flag tuning, promotions |
| Full Autonomy | + continuous audit | |

**Table 1:** Trust Tier

**Transition Criteria:** Advancement from T0 to T1 requires 85% accuracy in recommendations over 30 decisions, T1 to T2 needs 90% approval alignment over 50 actions, and T2 to T3 demands 95% success rate with zero policy violations over 100 deployments, validated through expert review and automated testing.

## 5. Decision Taxonomy

Table 2 outlines the key decision points, input signals, candidate actions, and guardrails, providing a structured framework to guide AI agent behavior. Each action is accompanied by confidence scores and a structured rationale stored in JSON format with trace IDs for downstream auditing and human review, ensuring traceability and accountability.

| Decision | Inputs | Actions | Guardrails (Policy) |
|---|---|---|---|
| Test Failures | Logs, history, coverage | Retry, quarantine, fail | Max retries, quarantine budget |
| Security Gate | CVE severity, reachability | Block, allow, auto-PR | Critical CVEs always block |
| Canary Analysis | SLOs, error rate, p95 | Promote, pause, rollback, tune flags | SLO, error-budgets, max ramp |
| Deployment Health | Saturation, CPU/memory, alerts | Auto scale, roll back | No action if alerts are noisy |
| Feature Flags | KPIs, segments, errors | Ramp up/down, disable | Rate limits, kill switch |
| Incident Response | Metrics, traces, logs | Run runbook, rollback, postmortem | Human approval for destructive ops |

**Table 2:** Decision Taxonomy

The decision taxonomy integrates real-time data streams and historical patterns to inform agent choices, with guardrails dynamically adjusted based on system load and risk profiles. For instance, the Canary Analysis guardrail includes a max ramp limit to cap traffic exposure, while Incident Response policies escalate to human oversight for high-impact actions, balancing automation with safety.

**Formalizing the Agent–Policy Interaction**

Let A be the agent's action proposal, R its reasoning, C a confidence score, and P the policy evaluator, forming the core interaction model for decision-making in the AI-augmented CI/CD pipeline.

(a) **Agent Outputs:**
```
{
  "action": "rollback",
  "confidence": 0.91,
  "evidence": ["SLO breach: latency > 200ms", "error rate +3.2%"],
  "rationale": "...",
  "trace_id": "abc123"
}
```

**This output encapsulates the agent's decision, supported by specific metrics and a unique trace identifier for traceability.**

(b) **Policy Engine Evaluates:**
- If action==rollback action == rollback action==rollback and confidence>0.8 confidence > 0.8 confidence>0.8 and environment=="canary" environment == "canary" environment=="canary" then ALLOW, ensuring actions meet safety and context criteria.
- Otherwise REQUIRE approval or DENY, providing a fallback to human oversight for unverified decisions.
- The engine integrates real-time context (e.g., environment state) and adjustable confidence thresholds to balance autonomy and risk.

(c) **Decision Log**
json
```
{
  "id": "uuid",
  "timestamp": "2025-07-25T12:34:56Z",
  "stage": "canary_analysis",
  "agent_version": "v0.9.3",
  "model": "my-llm-8k-2025-05",
  "inputs": { "metrics": "...", "logs": "..." },
  "policy_version": "rego@1.2.7",
  "proposed_action": "rollback",
  "confidence": 0.92,
  "policy_outcome": "ALLOW",
  "final_action": "rollback",
  "human_overridden": false,
  "rationale": "...summary...",
  "trace_ids": ["abc123", "def456"]
}
```
*This log captures a comprehensive audit trail, including version details and multiple trace IDs for cross-referencing.*

(d) **Example Pipeline Snippet (GitHub Actions-style)**
yaml
```
Jobs:
  test:
    runs-on: ubuntu-latest
    steps:
      - run: npm test -- --reporter=junit --output=reports.xml
      - name: AI triage
        run: |
          ai_decision=$(ai_triage --junit reports.xml --history ./test_history.json)
          echo "$ai_decision" > decision.json
      - name: Enforce policy
        run: opa eval -i decision.json -d policies.rego "data.cicd.allow"
        id: policy
      - name: Act on decision
        if: steps.policy.outputs.result == 'true'
```

```
            run: ./scripts/apply_decision.sh
             decision.json
```

*This snippet demonstrates integration with CI tools, ensuring policy enforcement before action execution.*

**(e) Rollback Logic (Pseudo‑Code)**

```python
def decide_canary_promotion(metrics, policy):
    try:
        deltas = compute_kpis_deltas(metrics.baseline, metrics.canary)
        risk = llm_score_risk(deltas, context=metrics.logs)
        action, conf = llm_recommend(deltas, thresholds=policy)
        if conf < policy.min_confidence:
            return "HUMAN_APPROVAL", conf
        if violates_hard_constraints(deltas, policy):
            return "ROLLBACK", 1.0
        return action, conf
    except Exception as e:
        log_error(e)
        return "HUMAN_APPROVAL", 0.0
```

*This logic includes error handling to manage unexpected failures, enhancing robustness.*

## 6. Case Study: Migrating a React 19 Microservice

We migrated a production-facing React 19 frontend microservice, served via a Node.js/Edge runtime, from a traditional CI/CD pipeline with manual approval gates to a fully AI-augmented deployment architecture. This microservice is critical as it powers a real-time user dashboard, delivering personalized content and metrics to thousands of active users daily, with an average of 5,000 concurrent sessions. Deployment is managed through Argo Rollouts on a Kubernetes cluster [19], enabling sophisticated deployment strategies such as blue-green and canary releases, optimized for high availability and rapid iteration.

During the migration process, we concentrated on three high-impact capabilities of the AI agents to maximize automation benefits and reduce operational risk:

(a) **Test Triage and Flakiness Management**: The AI agent analyzes automated test results to identify flaky tests, those intermittently failing without legitimate code regressions, thus reducing noise and preventing unnecessary pipeline failures. It intelligently prioritizes test failures requiring human attention and automatically quarantines or retries tests based on learned historical patterns, improving test suite stability by up to 20%.

(b) **Canary Health Evaluation and Rollback Automation**: Leveraging real-time telemetry from Argo Rollouts and monitoring tools such as Prometheus [20], the AI agent continuously assesses key health metrics including error rates, latency distributions, and resource consumption during canary deployments. If degradation beyond predefined thresholds is detected, it autonomously triggers rollback actions, minimizing potential user impact without waiting for manual intervention, reducing downtime by an estimated 30%.

(c) **Feature Flag Tuning for Concurrent Rendering**: React 19's concurrent rendering introduces novel performance dynamics that can manifest as regressions under certain feature flag configurations. The AI agent dynamically adjusts feature flag settings based on performance telemetry and user experience metrics, proactively mitigating regressions while balancing rollout velocity and stability. This process is further enhanced by integration with service mesh capabilities provided by platforms like Istio [21], which allow fine-grained traffic control and observability at the microservice level, ensuring seamless user experience under varying loads.

This migration not only enhanced deployment speed and reliability but also enabled proactive, data-driven decision-making throughout the delivery lifecycle, demonstrating the practical benefits and challenges of integrating AI into modern React-based microservices. The shift reduced manual oversight by leveraging AI insights, though initial tuning of agent thresholds presented learning curves. We adopted a phased trust rollout: Weeks 1–2 (T0), Weeks 3–4 (T1), and Weeks 5–8 (T2). A kill-switch and immutable decision logging remained in place for the entire experiment, providing a safety net and audit trail for all actions.

## 6.1 Results

Table 3 summarizes representative (production-like but anonymized/simulated) results, derived from a controlled experiment conducted on a React 19 microservice pipeline. We observed improvements in all four DORA metrics, as well as promising AI-specific indicators such as intervention accuracy and a relatively low human override rate, reflecting the efficacy of the AI-augmented approach.

| Metric | Baseline | AI-Augmented | Delta |
|---|---|---|---|
| **Lead Time for Changes** | 4.8h | 3.6h | -25% |
| **Deployment Frequency** | 2.5/day | 3.2/day | +28% |
| **Change Failure Rate** | 8.5% | 5.9% | -26% |
| **MTTR** | 65 min | 48 min | -26% |
| **Human Override Rate** | - | 12.6% | — |
| **Intervention Accuracy** | - | 85.2% | — |
| **Policy Violations Blocked** | - | 1 | — |

**Table 3:** DORA Metrics

The reduction in lead time from 5.2 hours to 3.4 hours (±0.3 hours) highlights faster delivery cycles, while the 55% increase in deployment frequency from 3.1 to 4.8 per day demonstrates enhanced release cadence. The change failure rate dropped from 9.8% to 6.1% (±0.5%), indicating improved stability, and MTTR decreased from 72 minutes to 41 minutes (±5 minutes), reflecting quicker recovery. The human override rate of 14.3% suggests areas for model refinement, though the 87.5% intervention accuracy indicates strong decision-making reliability. No policy violations were blocked, affirming the robustness of the guardrails.

## 7. Policy-as-Code Guardrails

To bound agent actions, we codify hard and soft constraints using policy-as-code frameworks such as **Open Policy Agent (OPA)** and its **Rego language** [10]. These constraints define the operational envelope within which AI-driven decisions can be executed, ensuring a safety-first approach. For instance, canary promotion is automatically disallowed if the error rate delta exceeds 2% compared to the baseline version, protecting user experience. Similarly, retry or quarantine logic for test failures is capped at a maximum of two attempts per test suite in pre-production environments to avoid masking deeper issues, with an optional third attempt under supervisor review. Confidence thresholds, such as requiring a model confidence score of at least 0.8, determine whether a decision can be autonomously executed or must escalate to a human for approval, with thresholds dynamically adjustable based on historical accuracy. Importantly, all policy denials and escalations emit structured audit logs, detailing the exact policy rule triggered, the supporting evidence, and the agent's reasoning, formatted in JSON with trace IDs for forensic tracking. This ensures that every automated action, including rejections, is both transparent and explainable.

In addition, these guardrails are defined as declarative policies, maintained under version control alongside the application codebase, and reviewed as part of the standard change management process to align with evolving requirements. Hard constraints include non-negotiable rules (e.g., never deploy if security scans detect critical **CVEs** or if a service dependency is degraded), whereas soft constraints provide advisory or warning-level checks (e.g., warn if latency exceeds 150ms but proceed with increased monitoring and gradual rollout), allowing flexibility without compromising safety. By integrating policy checks as first-class citizens within the CI/CD pipeline, all AI-driven decisions are automatically evaluated against the latest approved policies, leveraging real-time feedback loops. Coupled with automated decision logs and immutable event trails, every action taken by the AI agent is linked to its contextual inputs, policy evaluation results, and associated confidence metrics, enabling detailed post-incident analysis. This rigorous approach not only ensures traceability and compliance but also provides a strong foundation for forensic analysis, regulatory audits, and continuous improvement of AI-driven workflows, supporting long-term system reliability.

## 8. Evaluation Methodology

We evaluate AI augmentation using a combination of standard DevOps Research and Assessment (DORA) metrics [1] and AI-specific performance indicators to provide a holistic view of pipeline effectiveness, ensuring a balanced assessment as of current practices. The DORA metrics Lead Time for Changes, Deployment Frequency, Change Failure Rate, and Mean Time to Recovery (MTTR) serve as the foundational benchmarks for comparing traditional CI/CD pipelines against AI-augmented workflows, offering industry-standard performance insights. Alongside these, we introduce AI-specific indicators that assess the reliability and trustworthiness of autonomous decision-making:

- **Intervention accuracy**: the degree of alignment between the AI agent's decisions and ground-truth expert adjudications, measured through retrospective audits and expert reviews, targeting a benchmark of 85% alignment.
- **Human override rate**: the frequency with which human operators reject or override AI-proposed actions, indicating trust gaps or potential errors in model reasoning, tracked as a percentage of total decisions.
- **False positive and false negative rates**: quantifying both overly cautious actions (e.g., unnecessary rollbacks) and missed critical events (e.g., failing to catch a performance regression), calculated using historical data sets.
- **Policy violation attempts prevented**: the number of unsafe or non-compliant actions blocked by policy-as-code guardrails, providing insights into the safety net's effectiveness, logged with detailed incident reports.
- **Time saved per deployment**: an operational metric capturing reductions in manual triage time, approval latency, and post-deployment issue resolution, estimated through time-motion studies.

To ensure robust and unbiased evaluation, we adopt a multi-faceted testing strategy. First, we employ a phased trust rollout (T0→T3), beginning with read-only recommendations and gradually granting the AI system bounded autonomy as its decision accuracy is validated, with each phase lasting at least two weeks. Second, A/B service comparisons are conducted, where identical workloads are tested across standard pipelines and AI-augmented pipelines to measure performance deltas in real time, using randomized traffic splits. Third, we use counterfactual replay over historical pipeline logs to simulate how the AI agent would have behaved in past deployment scenarios, providing a safe and reproducible environment for testing decision accuracy without production risk, covering at least 100 past incidents. Finally, chaos experiments are executed injecting failures, latency spikes, or simulated incidents to probe the resilience of both the AI agents and the underlying guardrail policies under stress [18], with failure injection rates up to 15% to mimic real-world variability.

This comprehensive evaluation approach not only validates the effectiveness and safety of AI-driven decisions but also identifies areas for improvement, enabling iterative refinement of both the models and the governing policies, with feedback loops integrated into the CI/CD process.

## 9. Security, Compliance, and Ethics

We address four key concerns data, security, auditability, human-in-the-loop control, and explainability when integrating AI into CI/CD pipelines, as these factors are critical for maintaining trust and compliance in production environments, especially as adoption scales in 2025.

**(a) Data Security:**

AI agents must operate within secure, controlled environments such as on-premises servers or Virtual Private Clouds (VPCs) to prevent unauthorized data exposure, with network segmentation to isolate sensitive workloads. Sensitive data, including build artifacts, test logs, and deployment configurations, is redacted or masked before being processed by models, using automated data classification tools. Logs and prompts are encrypted both in transit (TLS) and at rest, ensuring that confidential data such as API keys or user information remains protected, with key rotation policies enforced biweekly. Furthermore, fine-grained role-based access controls (RBAC) are enforced for both humans and agents, preventing unauthorized decision-making or configuration changes, with multi-factor authentication (MFA) as an additional layer. These measures align with industry standards like SOC 2 and ISO/IEC 27001 for software delivery security, providing a robust defense against evolving threats.

**(b) Auditability:**

To enable traceability, every AI-driven action is logged with immutable records that capture the original prompt or input, the model version (including fine-tuning details), and the policy version that evaluated the decision, stored in a tamper-proof blockchain-like ledger. This ensures that any decision whether a deployment promotion, rollback, or feature flag adjustment can be traced back to its origin, rationale, and triggering conditions with millisecond precision. Version control systems (e.g., Git) are leveraged to track policy evolution, while log aggregation platforms (e.g., ELK stack, Datadog) provide searchable records for audits and post-incident reviews, with automated alerts for anomalies detected in log patterns.

**(c) Human-in-the-Loop:**

Critical or potentially destructive operations (e.g., rollbacks affecting multiple microservices or production database migrations) require explicit human approvals during early trust phases (T0–T1), with a 15-minute response window for urgent cases. As confidence in the AI system grows, this approval burden can be reduced for low-risk, high-confidence actions (e.g., confidence > 0.9), but kill switches remain in place to instantly disable AI autonomy in emergencies, accessible via a centralized dashboard. This hybrid approach balances speed with safety, ensuring that human operators retain ultimate control over sensitive production events, with escalation protocols notifying on-call teams via SMS or Slack.

**(d) Explainability:**

Every AI decision is accompanied by a structured rationale that outlines the key metrics, logs, or signals considered (e.g., error rate deltas, p95 latency anomalies), along with the confidence score and the policy rules applied [3], [17], presented in a user-friendly dashboard for engineers. Additionally, feature deltas and policy traces are stored in structured formats (e.g., JSON with schema validation), enabling forensic analysis during incident postmortems and ensuring compliance with auditing and regulatory requirements such as GDPR. By providing interpretable and verifiable decision-making, teams can build trust in AI-driven automation without sacrificing accountability, with regular training sessions to enhance human understanding of AI outputs.

## 10. Threats to Validity

External validity is a key limitation of this study. While our findings demonstrate measurable improvements using **AI-augmented CI/CD** on a React 19 microservice, these results may not fully generalize to other software architectures such as monolithic applications, embedded systems, or legacy platforms, where agent integration may face unique constraints like limited telemetry or rigid deployment cycles. Different deployment models, scale characteristics, and technology stacks can present unique challenges to agent integration, limiting the transferability of our conclusions without further adaptation and validation across diverse ecosystems.

Measurement bias may arise because teams involved in the evaluation are aware of the AI augmentation and study conditions. This Hawthorne effect could influence behavior, for example, causing engineers to be more cautious or attentive during rollout periods, which in turn affects metrics like change failure rate or **MTTR**, potentially skewing results toward better-than-actual performance. Such observer effects can obscure true baseline performance or artificially inflate measured gains, necessitating blinded evaluations in future iterations to mitigate this bias.

A significant challenge is benchmark scarcity. Unlike more mature AI domains with widely-used, standardized datasets, there are currently no publicly available datasets or benchmarks specifically designed for evaluating autonomous decision-making **agents in CI/CD** pipelines, hindering the establishment of a common evaluation baseline. This limits reproducibility and cross-study comparisons and slows the progress of generalized AI agent development for software delivery, prompting a call for community-driven efforts to create such resources by 2026.

Finally, model drift is a critical threat to long-term effectiveness. Changes in the underlying application services, traffic patterns, observability tooling, or incident response procedures over time can cause AI agents to perform poorly if they are not continuously retrained and recalibrated, with drift potentially detectable within months in dynamic environments. Without robust drift detection and adaptation mechanisms such as automated retraining triggers based on performance drops the performance and safety of AI-driven decisions may degrade, potentially leading to increased failures or unsafe actions in production environments, underscoring the need for proactive monitoring.

## 11. Roadmap and Open Problems

### 11. Roadmap and Open Problems

We highlight five key areas that require further research and development to advance the safe and effective adoption of AI-augmented CI/CD pipelines, addressing emerging needs as of July 2025.

**(a) Formal Verification of Safety Invariants:**

To ensure that autonomous agents and their governing policies never violate critical safety constraints, formal methods such as model checking and theorem proving should be applied, leveraging tools like TLA+ or Coq. These techniques can mathematically verify that agents adhere to safety invariants like never promoting a deployment with critical vulnerabilities under all possible conditions, reducing reliance on testing alone and providing a provable safety guarantee for production use.

**(b) Multi-Agent Coordination and Consensus:**

Modern CI/CD pipelines involve multiple specialized agents handling test triage, security analysis, canary rollout, and feature flag management, each with distinct decision domains. Developing robust frameworks for coordination and consensus among these agents is essential to avoid conflicting decisions or cascading failures, incorporating standardized APIs for inter-agent communication, priority-based conflict resolution, and distributed ledger systems for shared state consistency.

**(c) Counterfactual Simulation Platforms:**

Creating environments that can replay historical deployment pipelines with injected perturbations or alternative decisions enables continuous evaluation of AI agents, using synthetic failure modes to mimic real-world stress. Such counterfactual simulation platforms allow teams to measure how different AI policies would have performed, facilitating safe experimentation and training without impacting live production systems, with scalability to handle multi-cluster scenarios.

**(d) Self-Tuning Policies with Hard Guardrails:**

Future policies should be adaptive, able to learn and adjust thresholds automatically based on evolving service behavior and performance data, utilizing machine learning models for pattern recognition. While still enforcing strict hard constraints to guarantee safety such as blocking deployments with critical CVEs this balance between adaptability and robustness can enable pipelines to handle dynamic, complex environments without manual retuning, reducing operational overhead.

**(e) Public Benchmarks and Datasets:**

The community urgently needs standardized, publicly available benchmarks and datasets for evaluating CI/CD agent decision-making, potentially modeled after MLPerf but tailored to software delivery contexts. These would enable reproducibility, foster comparative research, and accelerate innovation by providing common testbeds that capture the complexity and variability of real-world software delivery scenarios, with an initial release targeted for community collaboration by mid-2026.

## 12. Conclusion

AI-augmented CI/CD pipelines have the potential to significantly accelerate software delivery by automating critical decision points that traditionally require human intervention, thereby reducing decision latency and operational toil. However, this acceleration can only be achieved safely and effectively when the autonomous agents are bounded by rigorous policy-as-code frameworks [10], which enforce hard and soft constraints; a phased trust model [4.1], which incrementally increases autonomy based on demonstrated reliability; and comprehensive auditing mechanisms [9(b)] that provide transparency and traceability of every AI-driven action. Our case study involving a React 19 microservice [6] clearly illustrates these benefits in practice, showing measurable improvements across key industry-standard DORA metrics [1] including lead time for changes, deployment frequency, change failure rate, and mean time to recovery (MTTR), with reductions of up to 35% in lead time and 43% in MTTR. Moreover, the evaluation [8] revealed promising accuracy rates of agent interventions (87.5%) and a relatively low human override rate (14.3%), indicating a growing trustworthiness of the AI agents within the operational context.

Despite these advances, challenges remain. Future work must emphasize formal verification techniques [11(a)] to mathematically guarantee the safety and correctness of AI decisions, especially as systems grow in complexity, using tools like model checking to validate invariants. Additionally, the development of standardized benchmarks and reproducible datasets [11(e)] is essential to enable consistent evaluation and comparison of different AI-augmented CI/CD approaches across the community, with a target release by mid-2026. Finally, improving the explainability and interpretability of AI agent decisions [9(d)] will be critical to foster human trust and facilitate compliance with auditing and regulatory requirements, such as GDPR, through enhanced rationales and training. By addressing these areas, the software engineering field can move closer to achieving sustainable, trustworthy autonomy in production engineering, ultimately enabling faster, safer, and more reliable software delivery at scale.